\documentclass[aps,prb,letterpaper,amsmath,amssymb,superscriptaddress,reprint]{revtex4-1}
\usepackage{graphicx}
\usepackage{microtype}
\usepackage{bm}
\usepackage{hyperref}

\begin{document}

\title{Large Spin Hall Effect in an Amorphous Binary Alloy}
\author{Markus Meinert}
\email{meinert@physik.uni-bielefeld.de}
\author{Katharina Fritz}
\affiliation{Center for Spinelectronic Materials and Devices, Department of Physics, Bielefeld University, D-33501 Bielefeld, Germany}
\author{Sebastian Wimmer}
\author{Hubert Ebert}
\affiliation{Department Chemie, Physikalische Chemie, Ludwig-Maximilians-Universit\"at M\"unchen, Germany}

\date{\today}

\begin{abstract}
We investigate the spin Hall effect of W-Hf thin films, which exhibit a phase transition from a segregated phase mixture to an amorphous alloy below 70\% W. The spin Hall angle was determined with a planar harmonic Hall voltage technique. Due to the accompanying jump in resistivity, the spin Hall angle shows a pronounced maximum at the composition of the phase transition. The spin Hall conductivity does, however, reduce from W to Hf with a weak discontinouity across the phase transition. The maximum spin Hall angle of $\theta_\mathrm{SH} = -0.25$ is obtained for amorphous W$_{0.7}$Hf$_{0.3}$. A detailed comparison with spin Hall conductivities calculated from first principles for hcp, fcc, and bcc solid solutions provides valuable insight into the alloying physics of this binary system.
\end{abstract}

\maketitle

\section{Introduction}

The spin Hall effect \cite{Dyakonov1971, Hirsch1999, Hoffmann2013, Sinova2015} converts a charge current $j$ into a transverse spin current $j_\mathrm{s}$. The charge-to-spin conversion efficiency is characterized by the spin Hall angle (SHA) $\theta_\mathrm{SH} = j_\mathrm{s} / j$. Effects from the band structure (intrinsic) as well as effects from extrinsic scattering (skew scattering and side-jump scattering) contribute to the spin Hall angle. In the dilute limit, the intrinsic and side-jump contributions depend on the resistivity, whereas the skew-scattering contribution is independent of the resistivity, i.e., $\theta_\mathrm{SH} = \tilde{\sigma}_\mathrm{SH} \rho_{xx} + b$, where $\tilde{\sigma}_\mathrm{SH}$ is the sum of intrinsic and side-jump spin Hall conductivity. Assuming the skew-scattering contribution to be negligible, i.e., for high impurity concentration, the SHA can be rewritten as $\theta_\mathrm{SH} = \tilde{\sigma}_\mathrm{SH} / \sigma_{xx}$. The spin Hall conductivity (SHC) of crystalline materials is both experimentally\cite{Qiu2013, Sagasta2016, Nguyen2016, Schulz2016, Zhang2017} and theoretically\cite{Tanaka2008, Lowitzer2011, Freimuth2010, Gradhand2012, Koedderitzsch2015} well understood and various metals with large spin Hall conductivity were identified, such as Pt\cite{Sagasta2016}, $\beta$-W \cite{Pai2012}, $\beta$-Ta \cite{Liu2012a}. The scaling relation between resistivity and spin Hall angle was experimentally verified for Pt thin films\cite{Sagasta2016}. This scaling also explains the large range of reported SHAs in the literature for a single material\cite{Sinova2015}. By doping Pt with Au it was shown\cite{Obstbaum2016} that the resistivity can be increased and that the spin Hall angle can be tuned by alloying. The spin current originating from the spin Hall effect can be absorbed by an adjacent ferromagnetic layer, where the angular momentum transfer gives rise to so-called spin-orbit torques \cite{Manchon2009}. These may induce precession of the magnetization \cite{Liu2012b}, domain wall motion \cite{Miron2010}, or switching of the magnetization orientation\cite{Miron2011, Pai2012, Liu2012a}. Various concepts for spin Hall based magnetic memory devices, so called spin-orbit torque magnetic random access memory (SOT-MRAM) were proposed\cite{Cubukcu2014, Garello2014, Fukami2016, Lau2016}, which hold promise for a more energy efficient and enduring memory cell as compared to spin-transfer torque (STT-)MRAM and even to conventional SRAM \cite{Prenat2016, Oboril2015}.

Tungsten is the element with the largest negative spin Hall conductivity\cite{Tanaka2008}. While the bcc phase of W has low resistivity and thus a small SHA, the $\beta$-phase of W typically has a high resistivity and thus an exceptionally high SHA up to -0.5, which was achieved by oxygen incorporation into the material \cite{Demasius2016}. However, due to the metastability of the $\beta$-W, this material is unsuitable for applications that require high-temperature annealing \cite{Neumann2016}. In the present work, we attempt to create a high-resistivity binary alloy thin film based on W by mixing with Hf. The binary phase diagram \cite{Lieser2012} contains a line compound at the stoichiometry W$_2$Hf with the cubic Laves structure (C15). Across the full range of composition, phase mixtures of bcc-W + W$_2$Hf or hcp-Hf + W$_2$Hf are expected with negligible mutual solubility at low temperature. We thus expect to obtain phase-segregated films with small grains and high resistivity and therefore large spin Hall angle. Instead, we discover an amorphous phase over a broad composition range which has both a high SHA and SHC.

\section{Methods}

\subsection{Experimental}
Thin films of W-Hf were grown by dc magnetron co-sputtering on thermally oxidized Si wafers at room temperature. The full stack was Si (001) / SiOx 50nm / $\mathrm{W_xHf_{1-x}}$ 8nm / Co$_{40}$Fe$_{40}$B$_{20}$ 3nm / TaOx 2nm. The growth rates of Hf and W were determined by a x-ray reflectivity to determine the power ratios of the W and Hf sources for the stoichiometry series. The Ar working pressure was $2 \times 10^{-3}\,\mathrm{mbar}$ and the base pressure of the deposition system was $5 \times 10^{-9}\,\mathrm{mbar}$. The alloy film thicknesses were confirmed by x-ray reflectivity. X-ray diffraction with Cu K$_\alpha$ radiation was performed in a diffractometer with Bragg-Brentano geometry. The film resistivities were determined by a four-probe technique with four equidistant needles in a line, such that the effective resistivity of the multilayer can be written as $\rho_{xx}^\mathrm{ML} = \frac{\pi}{\ln 2} \frac{U}{I} (t_\mathrm{W_xHf_{1-x}} + t_\mathrm{CFB})$. A parallel circuit model was subsequently applied with $\rho_\mathrm{CFB} = 175 \times 10^{-8}\,\Omega\mathrm{m}$ as determined for a 3nm CFB film to obtain the alloy resistivity $\rho_{xx}$.

For the determination of the spin Hall angle, the films were patterned into 4-fold rotationally symmetric Hall crosses with a conductor width of $w = 16\,\mathrm{\mu m}$ and a length of $l =  48\,\mathrm{\mu m}$ by optical lithography. Harmonic Hall voltage measurements were performed in a crossed Helmholtz coil setup with a maximum field of 20\,mT and the in-phase first harmonic and out-of-phase second harmonic Hall voltages were recorded simultaneously upon in-plane field rotation with a Zurich Instruments MFLI multi-demodulator lock-in amplifier. Additional measurements were performed for $x \geq 0.9$ in a dual Halbach cylinder array with a rotating magnetic field up to 0.5T (MultiMag, Magnetic Solutions Ltd.). The out-of-phase second harmonic voltage $V_\mathrm{2\omega}$ detected by the lock-in amplifier can be expressed as \cite{Wen2017}
\begin{equation}\label{eq:harmonichall}
V_{2\omega} = \left( \frac{- H_\mathrm{FL} \cos \varphi}{H - H_A} R_P \cos 2 \varphi + \frac{1}{2} \frac{H_\mathrm{DL} \cos \varphi}{H_K - H} R_A \right) I.
\end{equation}
The angle $\varphi$ is the angle between current and magnetization. The anisotropy field $\mu_0 H_K$ and the anomalous Hall resistance amplitude $R_A$ were obtained from a Hall voltage measurement in a perpendicular magnetic field up to 2.2\,T. The planar Hall amplitude $R_P$ is obtained from the first harmonic $V_\omega = R_P I \sin 2\varphi$. The in-plane anisotropy field $H_A$ is determined by finding the external field strength $H$ that saturates the planar Hall effect and gives rise to a sinusoidal response to the field rotation. The above formula was fitted to the experimental data and convergence of the effective field with respect to the external magnetic field strength was verified (see Appendix for details). The spin Hall angle was obtained from the damping-like effective field as
\begin{equation}
\theta_\mathrm{SH} = \frac{2e}{\hbar} \frac{ \mu_0 H_\mathrm{DL} M_\mathrm{s} t_\mathrm{CFB} }{j}.
\end{equation}
The magnetization of the CoFeB film was determined by alternating gradient magnetometry to be $M_\mathrm{s} = (1050 \pm 50)\,\mathrm{kA/m}$. In all cases, the field-like effective field is very small compared to the damping-like effective field and is therefore neglected.

\begin{figure}[t]
\includegraphics[width=8.6cm]{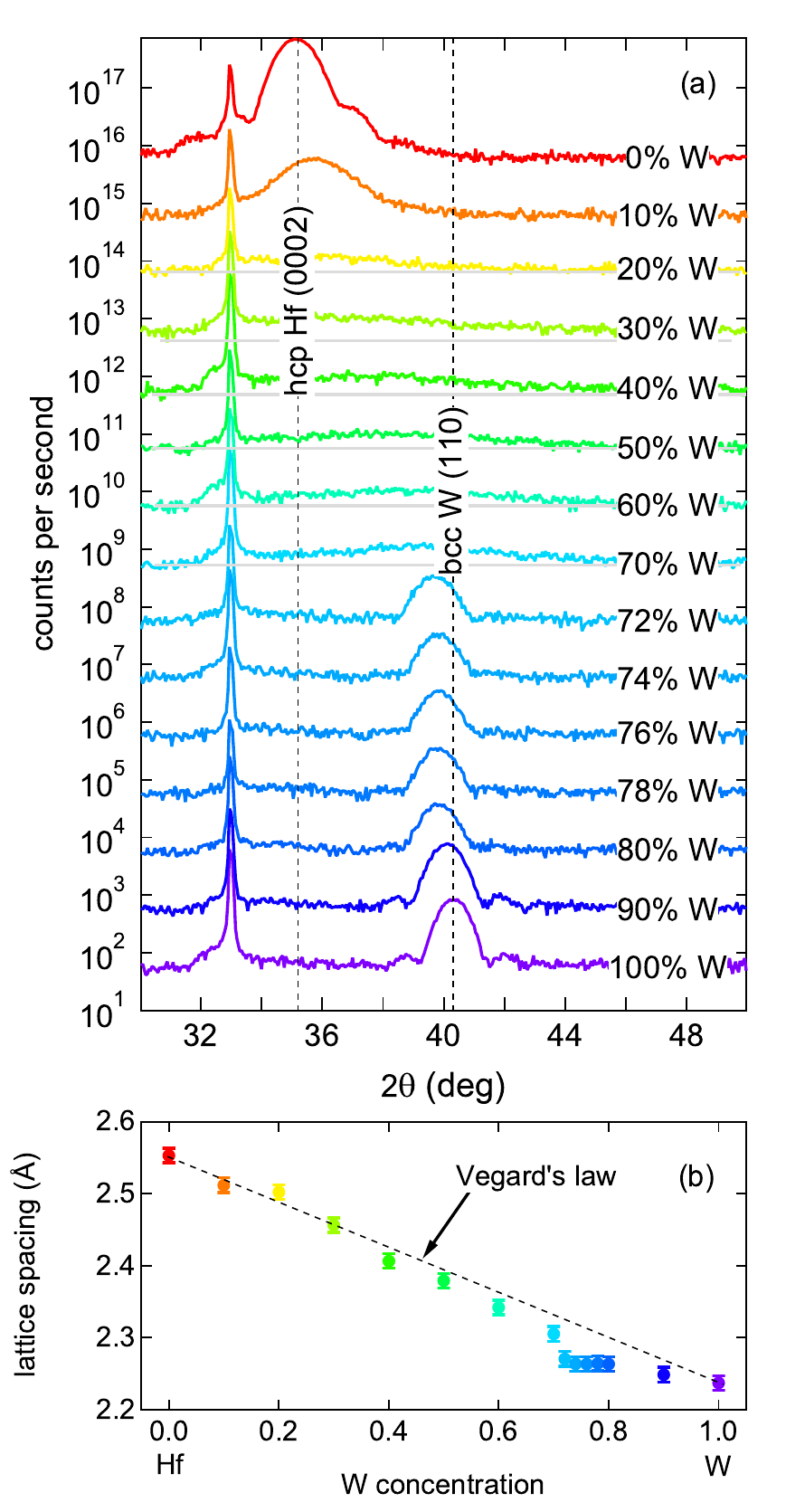}
\caption{\label{fig:xrd} Top: X-ray diffraction patterns of W$_x$Hf$_{1-x}$ 8nm / CoFeB 3nm / TaOx 2nm films. Bottom: Lattice spacing obtained from the diffraction peaks. Peaks in the composition range $x=0.7 \dots 0.2$ are weak and broad. Light grey lines are drawn to indicate the background levels and improve the visibility of the amorphous diffraction humps. The color code for the W-Hf stoichiometries is used throughout the article.}
\end{figure}

\subsection{First-Principles Calculations}
The spin Hall conductivities were calculated in a fully relativistic multiple-scattering Green function framework using a Kubo-Bastin formalism \cite{Lowitzer2011}. Intrinsic and extrinsic contributions to the spin Hall conductivity are treated on equal footing. Furthermore, chemical alloying as well as temperature are treated on equal footing within the coherent potential approximation (CPA), or the alloy-analogy model (AAM), respectively \cite{Ebert2015}. The formalism is implemented in the Munich Spin-Polarized Relativistic Korringa-Kohn-Rostoker (SPR-KKR) code \cite{Ebert2011, SPRKKR}. The Green function was expanded up to $\ell_\mathrm{max} = 3$ and the Fermi energy was accurately obtained with Lloyd's formula. The atomic sphere approximation (ASA) was used throughout. For the evaluation of the Kubo-Bastin formula 32 points were used for the energy integration. Approximately $5 \times 10^7$ k-points in the full Brillouin zone were used to ensure an accurate evaluation of the Brillouin zone integrals for the Fermi surface term. The spin Hall conductivities were calculated for bcc, fcc, and hcp solid solutions, where the atomic volumes and the Debye temperatures are interpolated between the experimental values according to Vegard's rule. Additionally, the spin Hall conductivity of W$_2$Hf (C15 Laves phase) was calculated using experimental lattice constants. The spin Hall conductivities were calculated at 300\,K in all cases.

\section{Structural characterization}

In Fig.\ref{fig:xrd}\,(a) we show x-ray diffraction patterns of our stoichiometry series. In the W-rich portion down to 72\% W,  a diffraction peak is found that can be indexed as bcc-W (110), in agreement with the expectation that bcc-W will grow with a strong (110) preferred growth direction on SiOx to minimize its interface energy. The strong diffraction peak in the pure Hf can be indexed as hcp-Hf (0002), again in agreement with the expectation that hcp-Hf will grow with a strong (0001) texture on SiOx. As Hf is added to W, the diffraction peaks shift to smaller angles and become slightly broader. The length of coherent scattering along the growth direction $D_z$ can be determined with Scherrer's formula, $D_z = k \lambda / (B \cos \theta)$, with $k=0.9$ and $B$ the observed peak full-width-half-maximum. At 100\% W, we obtain $D_z \approx 8\,\mathrm{nm}$, i.e., the film has a fibre texture with grains extending along the full thickness of the film. Additional evidence for the very good crystal quality comes from the presence of symmetric Laue oscillations around the diffraction peak. With increasing Hf content, the peaks become slightly broader and the Laue oscillations vanish. This indicates either a smaller $D_z$ or the presence of microstrain in the grains, i.e., a change of lattice constant along the growth direction.

In the composition range of $x=0.2 \dots 0.7$, broad humps are observed, with the position of the humps varying linearly with the stoichiometry. Using Scherrer's formula, $D_z \approx (1\pm 0.2)\,\mathrm{nm}$ is found in this regime. We interpret this as an amorphous phase that has some local order spanning three to four interatomic distances. The observed humps can be interpreted as local atomic arrangements of either bcc (110)-type, hcp (0001)-type, or fcc (111)-type.

In the W-rich part of the series, the interatomic distance $d_{hkl}$ (Fig. \ref{fig:xrd}\,(b)) does clearly deviate from Vegard's law, according to which it would vary linearly between the lattice spacings of bcc-W and bcc-Hf as a function of the stoichiometry parameter $x$. In fact, a saturation of the lattice spacing at $2.264\,\mathrm{\AA{}}$ is observed. This agrees with the binary phase diagram, according to which a mixture of bcc-W and W$_2$Hf is expected in this regime and essentially no Hf would dissolve in the bcc-W lattice. The slight expansion of the lattice spacing may then be due to a small fraction of Hf dissolved in bcc-W or due to compressive strain caused by Hf or W$_2$Hf precipitates in the grain boundaries. In either case, we conclude that for $x > 0.7$ no solid solution is formed, wheras the amorphous phase in the range $0.3 \leq x \leq 0.7$ is a homogenous solution of Hf and W, where the interatomic distance follows Vegard's law. We found no evidence for the formation of crystalline W$_2$Hf in the thin films.

\begin{figure}[t]
\includegraphics[width=8.6cm]{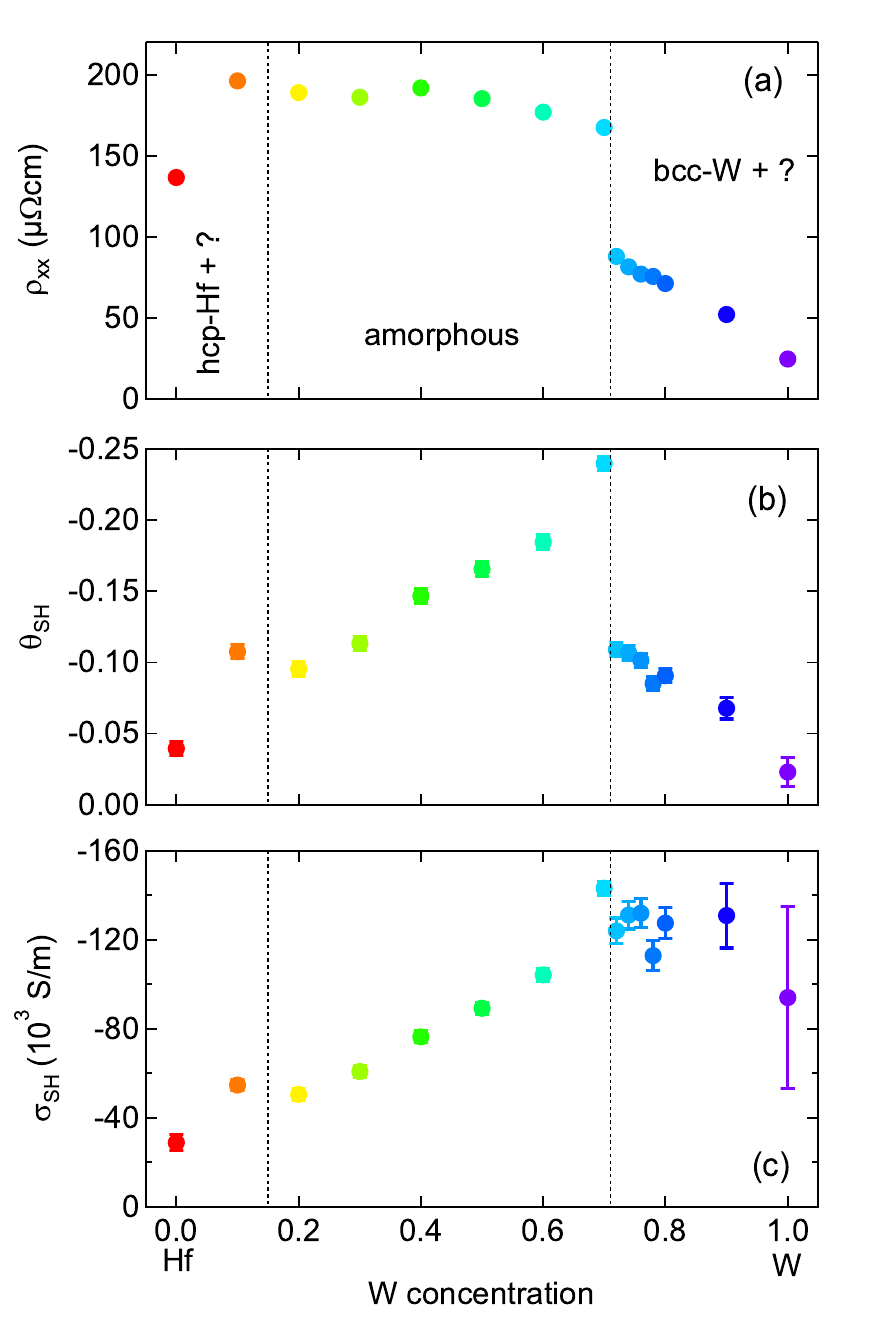}
\caption{\label{fig:elec} (a) Resistivity $\rho_{xx}$ of the 8nm W$_{x}$Hf$_{1-x}$ films. (b) Spin Hall angle determined with the planar Harmonic Hall method. (c) Spin Hall conductivity $\sigma_\mathrm{SH} = \theta_\mathrm{SH} / \rho_{xx}$. The dashed lines indicate the phase transformations observed in x-ray diffraction.}
\end{figure}

\section{Electrical Resistivity and Spin Hall Conductivity}

The electrical resistivity shows clear signs of the phase transition, as is shown in Fig. \ref{fig:elec} (a). Upon Hf addition into W, the resistivity is increased linearly, similar to what is expected for a solid solution. At the stoichiometry of the transition, the resistivity jumps by 90\%. In the amorphous regime, the resistivity increases up to W$_{0.1}$Hf$_{0.9}$. 

As shown in Figure \ref{fig:elec} (b), the spin Hall angle obtained from the planar harmonic Hall measurements resembles the increase of the resistivity in the W-rich films and exhibits a jump at the phase transition. The maximum spin Hall angle of $\theta_\mathrm{SH} = -0.25$ is obtained at $x = 0.7$. The spin Hall conductivity $\sigma_\mathrm{SH} = \theta_\mathrm{SH} / \rho_{xx}$ shows a nearly linear dependence on the composition irrespective of the phase transition, where only a small discontinuity is seen. The spin Hall conductivity (SHC) of the pure W film is difficult to obtain accurately with CoFeB as the ferromagnet, see the Appendix for details. We find $(0.94 \pm 0.41) \times 10^5$\,S/m, in fair agreement with a theoretical prediction \cite{Seifert2016}. The SHC at $x = 0.7$ is found as $1.43 \times 10^5$\,S/m. In the amorphous regime, a nearly linear decrease of the SHC with increasing Hf content is observed. The SHC of pure Hf is found to be $0.29 \times 10^5$\,S/m. Because of the rather thick films in our experiment, we do not correct for incomplete saturation of the diffusive spin current and assume that the film thickness is well beyond the spin-diffusion length across the whole composition series. We estimate the electron mean-free path $\lambda_0$ based on a detailed analysis of the resistivity of epitaxial W films in \onlinecite{Choi2012}, for which $\rho_0 \lambda_0 = 1.01 \times 10^{-15} \Omega \mathrm{m}^2$ was obtained, with the bulk resistivity $\rho_0$. In the amorphous phase, $\lambda_0 \approx 5.9 \times 10^{-10}\,\mathrm{m}$ is obtained, i.e., of the order of two interatomic distances, corroborating the amorphous character of the material. The spin-diffusion length $\lambda_\mathrm{sf}$ can be expressed with the spin-flip probability $p_\mathrm{sf}$ as $\lambda_\mathrm{sf} = \lambda_0/ p_\mathrm{sf} $. In heavy elements, the spin-flip probability is of the order 0.5 ($p_\mathrm{sf} = 0.57$ in Pt \cite{Sagasta2016}), so that the spin-diffusion length is estimated to be of the order of $\lambda_\mathrm{sf} \approx 1\,\mathrm{nm}$ in the amorphous regime. Therefore, the assumption of a saturated spin current is certainly justified in this case. However, strictly speaking, the spin Hall angles and conductivities given here are lower bounds to the true values in the respective bulk material.

\begin{figure}[t]
\includegraphics[width=8.6cm]{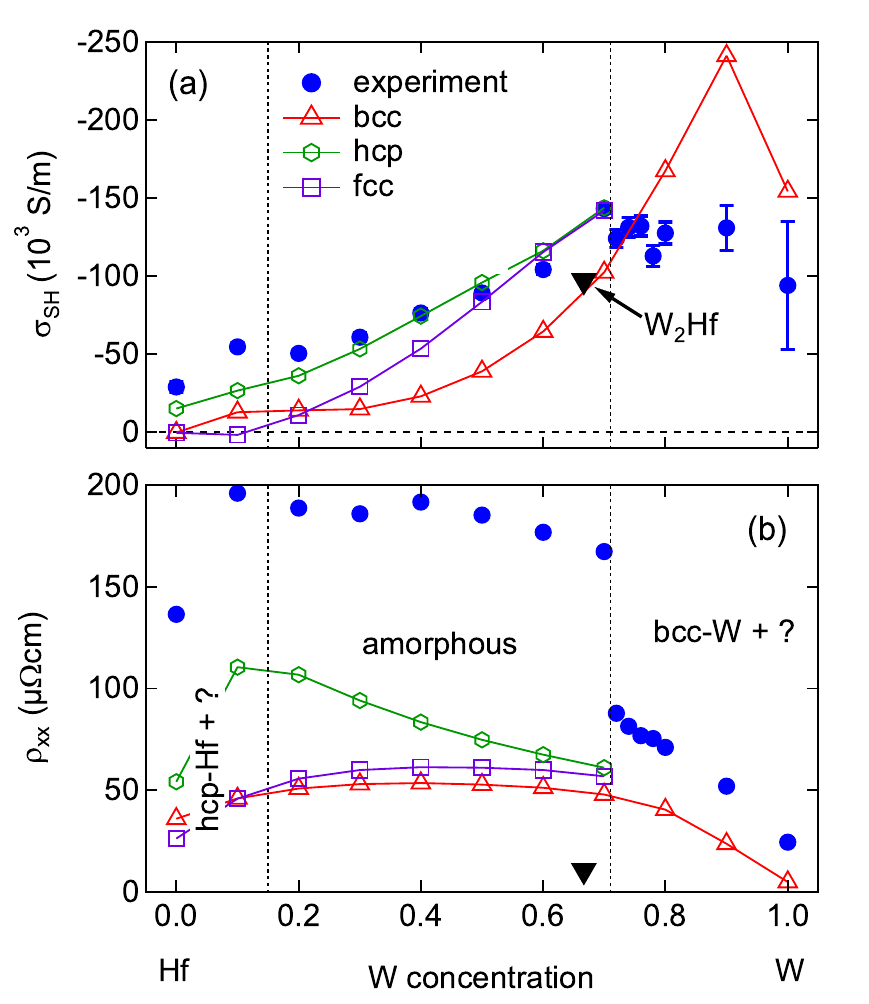}
\caption{\label{fig:kkr} Comparison of spin Hall conductivity (a) and resistivity (b), for experiment, three alloy models (bcc, hcp, fcc), and W$_2$Hf (C15) calculated at 300\,K.  The dashed lines indicate the phase transitions observed in x-ray diffraction.}
\end{figure}

To improve the understanding of the SHC in the W-Hf system, we compute it using the fully relativistic Kubo-Bastin formalism with the Munich SPR-KKR code. In Figure \ref{fig:kkr} we show a comparison of the experimental SHC and the calculated values for the three Bravais lattices considered. Results are shown including the so-called vertex corrections which however are negligible in the considered concentration range. Remarkably, the calculations for the bcc structure predict a sharp peak of the SHC at a composition of W$_{0.9}$Hf$_{0.1}$. This peak is absent in the experiment, in agreement with the inference from the lattice constant analysis that no solid solution is formed in the W-rich films. The calculated SHC is larger than the experimental value for pure bcc-W. In the previous section, we argued that two mechanisms can give rise to the peak-shift observed in the W-rich part of the series. Based on the SHC data, we rule out the possibility of large amounts of Hf dissolving in the bcc-W lattice, because in this case a strongly enhanced SHC should be observed. The approximately constant SHC in the W-rich regime can be understood by considering two counteracting effects: a small amount of Hf can dissolve in bcc-W, giving rise to an increase of the SHC, while the fraction of bcc-W reduces linearly with increased Hf content. Our results are thus consistent with less than 10\% Hf dissolved in W, while the remainder stays in the grain boundaries. The XRD peak shift is partially due to the lattice expansion by the Hf entering the bcc-W lattice and partially due to compressive strain acting on the bcc-W grains caused by the Hf segregation into the grain boundaries.

The amorphous part of the stoichiometry series is best described by the calculation for the hcp solid solution with (0001) texture (we plot $\sigma_{xz}^{y}$ in Fig. \ref{fig:kkr}(a)) and near-quantitative agreement between experiment and calculation is obtained. The fcc calculation deviates significantly at low W concentration, while the bcc calculation is consistently below the experimental data, in some cases up to a factor of 4. We therefore conclude that the local atomic arrangement in the amorphous phase is of the hcp type with a preferred (0001) orientation, which gives rise to the broad hump in the x-ray diffraction patterns. Furthermore, the slight increase of the electrical resistivity with increased Hf content and the sharp drop when approaching pure Hf is best reproduced by the hcp calculation (Fig. \ref{fig:kkr}(b)), wheras the fcc and bcc models predict a smooth and broad cusp for the resistivity with the maximum close to W$_{0.5}$Hf$_{0.5}$. Thus, we conclude that the amorphous phase has a local hcp-like atomic arrangement.

\section{Conclusion}

In summary, we obtained a large spin Hall angle up to $\theta_\mathrm{SH} = -0.25$ in an amorphous W-Hf phase, which has local order with a correlation length of about 1\,nm. We demonstrated that the spin Hall conductivity of the amorphous material can be understood in terms of the Kubo-Bastin formalism for periodic solids, making use of the local atomic arrangement. By comparison with calculations for the fcc, bcc, and hcp structures, we concluded that the local atomic arrangement in the amorphous W-Hf is of the hcp-type. We predict that a bcc solid solution of W$_{0.9}$Hf$_{0.1}$ has a particularly large spin Hall conductivity. However, in the experiment no such peak is observed, in agreement with the result that no homogenous solid solution is formed in sputtered W-Hf films at this composition. By means of a low-energy deposition technique such as molecular beam epitaxy (MBE) it might be possible to enforce the formation of a solid solution despite its positive formation enthalpy \cite{Lieser2012} and thus obtain a material with a very high negative spin Hall conductivity. Because amorphous materials contain no grains, these may become an experimental platform for pinning-free Skyrmion dynamics \cite{Lagrand2017}, where the Skyrmions are efficiently manipulated by the large spin Hall effect. This work demonstrates the utility of first principes calculations of the spin Hall conductivity for a rational materials design of novel intermetallic systems with large spin Hall angles.

\begin{acknowledgments}
The authors thank G. Reiss for making available the laboratory equipment. They further thank Tristan Matalla-Wagner for support with the high-field harmonic Hall measurements.
\end{acknowledgments}

\appendix*
\section{}

\begin{figure}[t]
\includegraphics[width=8.6cm]{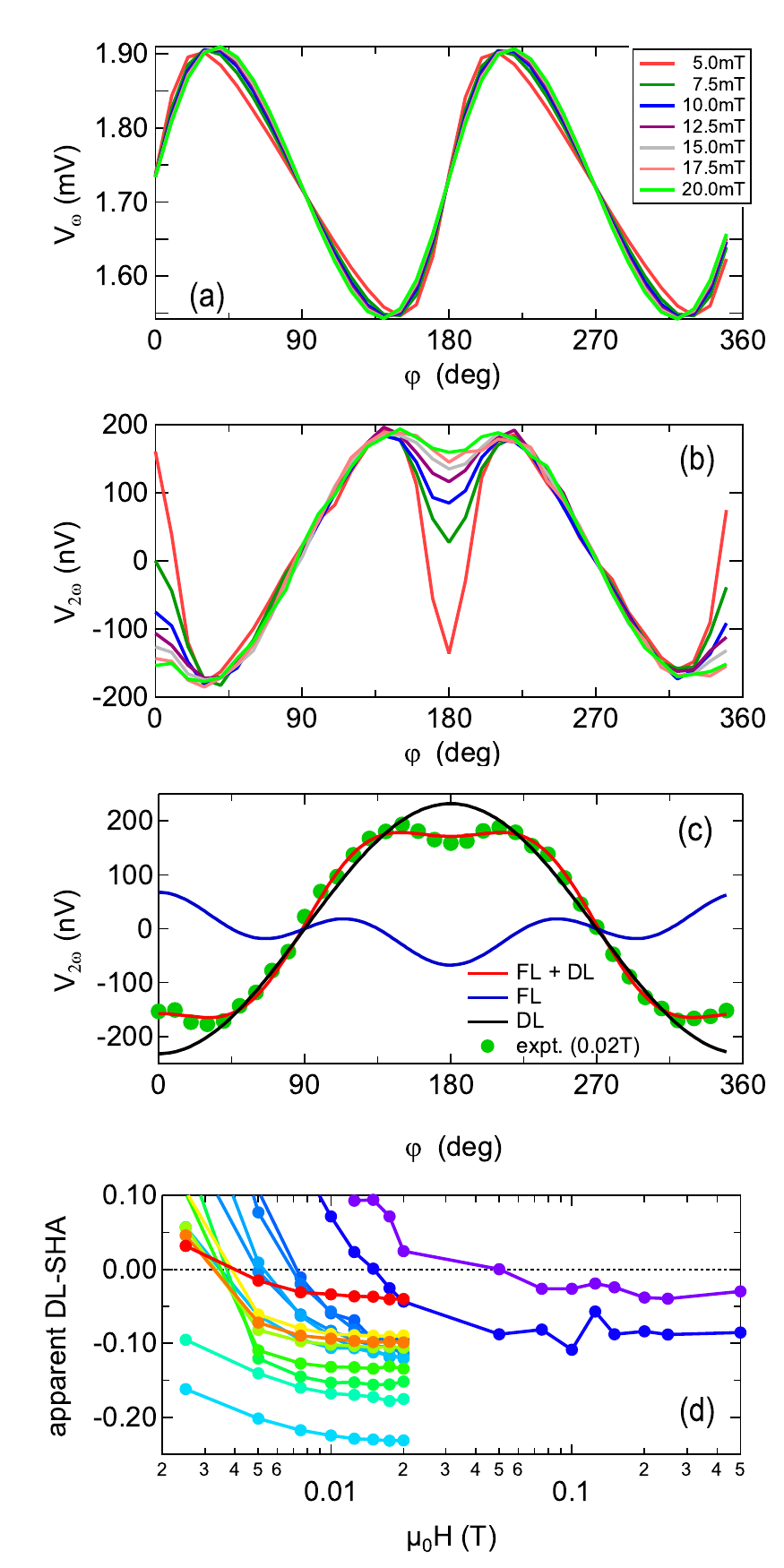}
\caption{\label{fig:example_analysis} (a): First harmonic in-phase signal of W$_{0.7}$Hf$_{0.3}$ for different in-plane magnetic fields. (b): Corresponding second harmonic out-of-phase signals. (c): Example analysis of the second harmonic Hall voltage and fit of Equation \ref{eq:harmonichall}. (d): Dependence of the apparent damping-like spin Hall angle on the in-plane magnetic field.}
\end{figure}

In Figure \ref{fig:example_analysis} we exemplarily show the dependence of the planar harmonic Hall analysis on the external field strength. The in-phase first harmonic signal $V_\omega$ (\ref{fig:example_analysis}) shows the usual $\sin 2\varphi$ behaviour. At small magnetic field, a two-fold anisotropy is superimposed on the signal, which points at a small uniaxial anisotropy in the films. At large magnetic fields, the anisotropy is suppressed and the magnetization follows the external magnetic field. 

The out-of-phase second harmonic signal $V_{2\omega}$ is composed of two contributions, the field-like (FL) and the damping-like (DL) term in Equation \ref{eq:harmonichall}. As predicted by the equation, the FL term vanishes approximately as $1/H$, whereas the DL term remains unaffected for $H \ll |H_K|$, see Figure \ref{fig:example_analysis}\,(b). By fitting Equation \ref{eq:harmonichall} to the data, the two contributions can be cleanly distinguished, as is shown in Figure \ref{fig:example_analysis}\,(c).

The uniaxial anisotropy gives rise to a deviation of the measured signal shape from the ideal behaviour predicted by Equation \ref{eq:harmonichall}, which results in erroneous values of the damping-like spin Hall angle. In Figure \ref{fig:example_analysis}\,(d), we show the apparent damping-like spin Hall angle as a function of the planar magnetic field. The spin hall angle converges to constant values in all high-resistivity samples for magnetic fields above 15\,mT. However, the low-resistivity samples with high W concentration require much larger magnetic field to achieve the convergence and measurements with magnetic fields up to 0.5\,T were performed. These measurements were additionally hindered by the small anomalous Hall resistance $R_A$, which arises from the shorting of the current by the low-resistivity W layer. To enhance the signal-to-noise ratio, larger current densities of up to $3.5 \times 10^{10}$\,A/m$^2$ were used to exploit that $V_{2\omega} \propto j^2$. Thus, it is advisable to use ferromagnetic films that have similar resistivity as the material to probe in harmonic Hall measurements. The spin Hall angles given in Figure \ref{fig:elec} were calculated as averages over several magnetic field strengths in the converged regime and results were averaged over at least three Hall crosses per sample.

\end{document}